\documentclass[iop]{emulateapj}
\usepackage{enumerate,color,pdflscape,lscape}

\newcommand{\HI}{H{\sc i}}

\shorttitle{Do Baryons Alter LSB Halos?}
\shortauthors{Kuzio de Naray \& Spekkens}

\begin{document}
\submitted{Accepted to ApJ Letters}
\title{Do Baryons Alter the Halos of Low Surface Brightness Galaxies?}
\author{Rachel Kuzio de Naray \& Kristine Spekkens}
\affil{Department of Physics, Royal Military College of Canada, P.O. Box 17000, Station Forces, Kingston, ON, K7K 7B4, Canada}
\email{kuzio@rmc.ca}
\email{Kristine.Spekkens@rmc.ca}

\begin{abstract}
High-quality observations of dark matter-dominated low surface brightness (LSB) galaxies indicate that, in contrast to the triaxial, centrally-concentrated cuspy halos formed in collisionless simulations of halo assembly, these galaxies reside in round, roughly constant density cored halos.  In order to reconcile these data with galaxy formation in the context of $\Lambda$CDM, processes that alter the shape and density  structure of the inner halo are required.  We compile observational properties of LSB galaxies to evaluate the plausibility that a previously higher baryonic mass content and feedback from star formation can modify the dark matter halos of these galaxies.   We also compare the properties of bulgeless disk galaxies formed in recent simulations to the LSB galaxy sample.   We find that observational constraints on LSB galaxy star formation histories, structure, and kinematics make it difficult for baryonic physics to sphericalize and decrease the central density of the dark matter halos of LSB galaxies.
\end{abstract}


\section{Introduction}
\label{intro}
There is now strong evidence that while galaxy formation is fundamentally driven by the hierarchical assembly of dark matter halos within the $\Lambda$CDM framework \citep[e.g.,][and references therein]{Navarro1996a, Stadel2009}, baryons play an important role in producing the observed structure of galaxies today.  The incorporation of baryons in simulations of galaxy formation has been instrumental, for example, in producing realistic disk galaxies that are only moderately bulged \citep[e.g.,][]{Brook2011, Guedes2011}.  Baryons have also been shown to be able to circularize the halo potential, transport angular momentum, and redistribute dark matter  (e.g., \citealp[][hereafter K10]{Kaz2010}; \citealp{Hopkins2009}; \citealp[][and references therein]{Weinberg2002}).

\defcitealias{Kaz2010}{K10}

It is therefore plausible that baryons alter the cuspy ($\rho \sim r^{-1}$ at small $r$), triaxial halos that are generically produced in collisionless simulations of halo assembly during galaxy formation.  Baryons are thus a double-edged sword:  while the discrepancies between collisionless simulations and the observed properties of galaxies provide important constraints on the baryonic physics that shapes galaxies, this same physics prevents a direct comparison between the predictions of collisionless simulations and the structure of halos inferred from observations for the majority of galaxies.  

There \textit{are} classes of galaxies, however, for which the comparison between collisionless theory and observations is thought to be more direct.  Conventional wisdom dictates that the halo structure of systems in which the dark matter dominates the mass density at all radii should resemble that produced by collisionless simulations.  Low mass (dwarf) and low surface brightness (LSB) galaxies are attractive candidates for comparison to $\Lambda$CDM predictions in this regard \citep[e.g.,][]{dBM97}.  

Despite disputes over the interpretation of early kinematic data, it is now clear that the halos inferred from observations of these systems are more core-like ($\rho \sim r^{0}$ at small $r$) and rounder (at least in the regions probed by the data) than $\Lambda$CDM predicts, creating a long-standing conflict between theory and observation in these galaxies \citep[see][and references therein]{deBlok2010}.  Recently, however, simulations of dwarf galaxies by \citet[][hereafter G10]{Governato2010} have used feedback from star formation to remove low angular momentum gas during galaxy formation, producing a dark matter-dominated present-day system, preventing the formation of a bulge, \textit{and} changing the initially cuspy dark matter halo into a more core-like halo.  It therefore seems that baryons are likely integral to resolving the apparent discrepancy between dwarf galaxy properties and $\Lambda$CDM predictions.  Are baryons equally influential during the formation and evolution of LSBs?

\defcitealias{Governato2010}{G10}

The kinematics, structure, and star formation histories of LSBs are now well-characterized.  In Table~\ref{datatable}, we compile a subset of the properties of a representative sample of blue, late-type, bulgeless LSBs  with central surface brightnesses fainter than $\mu_{0,B} \sim$ 23 mag arcsec$^{-2}$.  The table clearly illustrates that  a) LSBs span a range of masses that crosses the typical dwarf/high surface brightness (HSB) galaxy threshold, b) they are strongly dark matter-dominated with most of their baryons in \HI, and c) they are metal-poor with low past and present-day star formation rates.  It is important to note that LSBs are not rare; they comprise $\sim$ 50\% of the general galaxy population \citep[e.g.,][]{McGaugh1995}.

Cosmological galaxy formation simulations that aim to produce LSBs with cored, spherical halos must be able to match the basic properties in Table~\ref{datatable}.  While the \citetalias{Governato2010} simulations have utilized feedback to ease the conflict between collisionless halo predictions and dwarf galaxy observations, it remains unclear whether a similar mechanism applies to LSBs.

With the wealth of high-quality observations and improved galaxy formation simulations that are now available, the time is ripe to re-examine LSBs in the $\Lambda$CDM paradigm.  In this \textit{Letter}, we investigate the plausibility that currently favored baryonic processes \citepalias{Governato2010, Kaz2010} have altered the inner halos of LSBs in a way that reconciles the observed halo properties with collisionless simulation predictions.  We focus on two key questions: 1) can baryons sphericalize LSB galaxy halos, and 2) can baryons turn LSB galaxy halo cusps into cores?

\section{Sphericalizing the Halo} 
\label{kaz}
\citetalias{Kaz2010} have shown that galaxies with massive disks can modify the shapes of their dark matter halo potentials.  If the gravitational importance of the disk is significant enough, a triaxial halo can be sphericalized.  Quantitatively, \citetalias{Kaz2010} found that the halo can be affected by the disk if the disk contributes at least 50\% of the total rotation velocity at 2.2 times the disk scale length ($\eta \equiv V_{disk}/V_{circ} \,\gtrsim 0.5$ at $r = 2.2R_{disk}$).

\begin{figure}
\epsscale{1.2}
\plotone{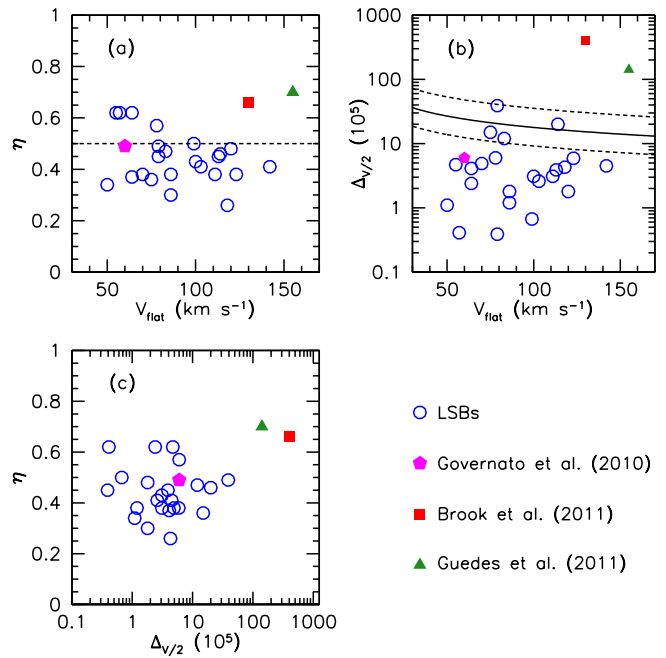}
\caption{(a): Gravitational significance of the baryonic disk.  Galaxies with $\eta > 0.5$ (dashed line) are able to influence the shape of their dark matter halos.  (b): The \citet{Alam2002} measure of central halo density.  The lines indicate the expected relation and scatter for NFW halos and WMAP5 $\Lambda$CDM parameters \citep{Maccio2008}. (c): Gravitational significance of the disk versus halo central density.  There is no correlation between these parameters for the LSBs.}
\label{dataplot}
\end{figure}

In Table~\ref{datatable}, we list the total, \HI, and stellar masses inside the last measured rotation curve point for the compiled LSBs, all of which have high-quality, two-dimensional H$\alpha$ and \HI\ kinematics (see Table references).  Also listed are the values of $\eta$ that we compute from the published mass models for each galaxy. By the \citetalias{Kaz2010} criterion,  only 4 galaxies clearly have disks massive enough to sphericalize the halo;  $\langle \eta \rangle \sim 0.41$ for the rest of the sample and there is no discernible trend with galaxy mass (see Figure~\ref{dataplot}a).

These subdominant LSB disks should therefore be embedded in triaxial dark matter halos.  However, constraints on the elongation of the halo potential from the harmonic decomposition of high-resolution velocity field data \citep[e.g.,][]{Trachternach2008} show the observations to be consistent with round potentials.  Similarly, model velocity fields and rotation curves in asymmetric potentials are inconsistent with observations: kinematic signatures of asymmetric potentials (e.g., misaligned kinematic and photometric axes, twisted kinematic minor axis) without photometric counterparts are largely absent from high-resolution velocity field data \citep{Kuzio2008,Kuzio2009,Kuzio2011}. The inner halos are round. 

 For comparison, we plot in Figure~\ref{dataplot}a the values of $\eta$ for three minimally-bulged disk galaxies formed in recent simulations.  The \citetalias{Governato2010} dwarf galaxy has just enough disk mass to sphericalize the dark matter halo. But the massive \citet{Brook2011} and \citet{Guedes2011} galaxies lie well above the sphericalization threshold, standing in stark contrast to LSBs of similar mass. 

Present-day LSB disks are not massive enough to have reshaped their dark matter halos.  In order to reconcile the observed round inner halos with collisionless triaxial halos using the \citetalias{Kaz2010} mechanism, LSBs must have had more baryons in the past; we return to this issue in Section~\ref{reconcile}.   

\section{Changing a Cusp to a Core}
\label{c2c}

As discussed in Section~\ref{intro}, LSB galaxy rotation curves are often more consistent with cored inner halos than the cusps produced by collisionless simulations. 
In Table~\ref{datatable}, we list the \citet{Alam2002} central halo density measure $\Delta_{V/2}$ --- the density within the radius at which the rotation curve falls to half its peak value, in units of the critical density --- as a proxy for the inner halo profile slope in our LSB sample.  We plot in Figure~\ref{dataplot}b the results for the LSBs and simulated galaxies, as well as the expected relation for NFW halos \citep{Navarro1996a} in the WMAP5 $\Lambda$CDM cosmology \citep{Maccio2008}.  For the LSBs, we find $\langle \Delta_{V/2} \rangle \sim 6 \times 10^5$, approximately a factor of 4 lower than expected.  To reconcile this number with $\Lambda$CDM, LSBs must have been formed in cuspy halos that were altered by some physical process that produces no observable trends between $\Delta_{V/2}$ and either $V_{flat}$ or $\eta$ (see Figure~\ref{dataplot}). 

A number of ways of changing cuspy halos into cored halos have been suggested (e.g., dynamical friction, disk bars, supernova feedback), and generally invoke baryons.  Feedback from star formation is of particular interest in the context of dark matter-dominated, bulgeless systems, since simulations suggest it can be effective at transforming cusps into cores (\citealp{Navarro1996b}; \citetalias{Governato2010}). Additionally, feedback in the \citetalias{Governato2010} simulations produce $\Delta_{V/2}$ and $\eta$ that are similar to LSB values. This feedback and outflow process can either be strong, impulsive and relatively violent, or as more recently suggested, can take place through a series of small(er) star formation events at $z > 1$ \citep[][hereafter P11]{Pontzen2011}.      
  
Invoking feedback to explain LSB galaxy halo cores therefore implies that the galaxies have had at least one period of early star formation and baryonic mass loss. This seems at odds with their low star formation rates, low surface densities and blue colors (see Table~\ref{datatable}). We explore possibilities for reconciling LSB galaxy properties with the feedback mechanism for producing cores in Section \ref{reconcile}.

\defcitealias{Pontzen2011}{P11}

\section{Understanding LSB Properties: The Role of Mass Surface Density}
\label{surfdensity}
Low mass surface densities are likely key to many of the observed properties of LSBs.  Star formation is both minimal and inefficient when the gas density is low.  A galaxy that forms few(er) stars naturally has low metallicity, low surface brightness, a large gas mass fraction, and a large mass-to-light ratio, all properties describing the systems in Table \ref{datatable}.

One way to achieve low gas density in LSBs is to form them in underdense regions \citep[e.g.,][]{Rosenbaum2009}. External processes like tidal interactions and mergers that increase the gas density are minimized if the overdensities that eventually become LSBs were originally located in or near large-scale voids.  This is an attractive formation scenario for LSBs as it would also satisfy the observational constraints that the galaxies are relatively isolated and located on the edges of large scale structure \citep{Bothun1997,Rosenbaum2009}.  

The gas densities may also be kept low if LSBs form within high spin parameter halos \citep[e.g.,][]{Dalcanton1997,Boissier2003}.  High angular momentum suppresses the collapse of the disk, leading to a galaxy with a larger disk size, lower surface brightness, and lower gas surface density than a low-spin galaxy of the same mass.  Additionally, these galaxies will have larger dynamical mass-to-light ratios and rotation curves that are determined by the dark matter rather than the baryons.  All of these characteristics apply to LSBs \citep[e.g.][]{Zwaan1995}.

The challenge for LSB galaxy formation within $\Lambda$CDM is to alter the halos, presumably via a baryonic process, in a way that preserves or produces the low present-day surface densities that explain the properties in Table~\ref{datatable} and Figure~\ref{dataplot}.

\section{Reconciling LSB Halos with $\Lambda$CDM Predictions}
\label{reconcile}
The work of \citetalias{Kaz2010} has demonstrated that massive baryonic disks can sphericalize inner halos, and that of \citetalias{Governato2010} and \citetalias{Pontzen2011} suggest that star formation feedback can flatten an inner density cusp in dwarf galaxies.   Extending these ideas to the LSB regime, it may be possible to understand present-day LSB galaxy halo properties in the $\Lambda$CDM context if these systems had substantially more baryons in the past that were then blown out after star formation. 

In the \citetalias{Governato2010} simulations, ejecting ``a few times the current stellar mass" is sufficient to reproduce observed dwarf galaxy properties after the halo core is formed.  For the LSBs in Table~\ref{datatable}, this conservatively translates to a blown-out baryonic mass of $\sim 6 \times 10^{9} M_{\sun}$. The value of $\eta$ before the mass loss is presumably higher as well, easing the tension between the present-day roundness of the inner halos and the \citetalias{Kaz2010} criterion.

Here, we discuss the plausibility of this model for round, cored halos in LSBs given their observed properties in Table~\ref{datatable}.  Specifically, we address how star formation would be induced (Section~\ref{starform}), the effectiveness of the subsequent feedback (Section~\ref{feedback}), and whether early and/or multiple bursts of star formation are consistent with the star formation histories of LSBs (Section~\ref{history}).

\subsection{Inducing Star Formation}
\label{starform}
Mergers and interactions are effective ways to generate star formation by funneling gas to the galaxy center.  This may be a tall order for LSBs considering that they have fewer, and more distant, neighbors than HSBs  \citep{Bothun1997,Rosenbaum2009}. This mechanism is even more problematic if LSBs form in underdense regions and then migrate to their current locations (see Section~\ref{surfdensity}). Observations support a quiescent evolutionary picture in which LSBs have had fewer major mergers and tidal interactions than HSBs.  

Secular processes like bar and bulge formation are also effective at directing low angular momentum gas to the centers of galaxies where star formation can occur. These structures are largely absent from LSBs, however \citep{Bothun1997}; this is not surprising given that light disks are stable against bar formation \citep[e.g.,][]{Mihos1997}.  

There is an additional hurdle to initiating star formation in LSBs if they form in high-spin halos (see Section~\ref{surfdensity}).  If this is the case, then LSBs are already at a disadvantage:  there would be an even higher angular momentum threshold to overcome. 
 
Initiating starbursts in LSBs therefore seems difficult. Nevertheless, if one manages to induce efficient star formation at some point in their evolution, how effective is the resulting feedback?

\subsection{The Effectiveness of Feedback}
\label{feedback}
The values of $\eta$ in Table~\ref{datatable} suggest that baryons must be removed from LSBs after their halos are sphericalized.  For the lower mass LSBs, this may not be problematic.  The \citetalias{Governato2010} simulations have shown that star formation feedback can eject significant amounts of baryons from dwarf galaxies, a process that is in fact required to match observations \citepalias[see also][]{Pontzen2011}. 

However, LSBs are not exclusively low mass (see Table~\ref{datatable}), and supernova winds are not expected to be effective at ejecting baryons from systems with $V_{flat} \gtrsim 100$ km s$^{-1}$ \citep[e.g.,][]{Dekel1986}.  Similarly, as discussed by \citet{Keres2009} and references therein, unless the star formation rate is very high (as at high redshift), feedback and outflows do not eject large amounts of baryons in galaxies with $M_{galaxy} > 10^{9} M_{\sun}$.  This is the realm in which the LSBs we are considering fall; their gas masses alone are a few~$\times 10^{9} M_{\sun}$ (see Table \ref{datatable}).   \citet{Brook2011} have shown that baryons are only temporarily blown out of high mass galaxies and eventually fall back to the galaxy disk and form stars.  This scenario would be inconsistent with the $\eta$ values determined for higher mass LSBs today.  

For a feedback scenario to be successful in LSBs, it must be effective at removing baryons from LSBs with a range of masses.  Notwithstanding this challenge, if one assumes that enough star formation and feedback occur to alter LSB galaxy halos and blow out baryons, is the resulting stellar population consistent with the observed characteristics of LSBs today?

\subsection{LSB Galaxy Star Formation Histories}
\label{history}
Observations of the stellar populations, star formation histories, colors, and metallicities of LSBs place further constraints on the likelihood that baryons alter the dark matter halos and are subsequently ejected by star formation and feedback. 

Attempts to constrain the ages of LSBs find stellar populations in the range of 1.5\,--\,6 Gyr \citep[][and references therein]{Vorobyov2009}, suggesting that the major star formation event took place at a redshift between $z \sim$ 0.2\,--\,0.4 \citep{Haberzettl2008}.  Substantial populations of old stars, which would exist if large-scale star formation had taken place in the past, are not observed; young stars, not a dominant population of old, low mass stars, are responsible for the blue colors of LSBs (see Table~\ref{datatable}).  LSBs are not the faded remnants of once-HSBs that have ceased star formation \citep[e.g.,][]{Wyder2009}.

LSBs are also metal-poor (see Table~\ref{datatable}), indicating that they form relatively few stars over a Hubble time.  Additionally, there is typically not much dust, a byproduct of star formation, in LSBs \citep[see the discussion and references in][]{Wyder2009}.

Current star formation rates in LSBs are at least an order of magnitude lower than in HSBs \citep[][see also Table~\ref{datatable}]{vdHoek2000}, and the star formation efficiencies are only a few percent \citep[e.g.,][]{Wyder2009}.   Observations and modeling indicate that the properties of LSBs are in good agreement with exponentially decreasing star formation rates combined with sporadic small-amplitude star formation events \citep[e.g.][]{vdHoek2000, Vorobyov2009}.  

Given the observational constraints of gas-richness, low gas surface density, low star formation rates, and low abundances, the amount of past star formation in these galaxies could not have been very large; this is borne out by estimates of their past star formation rates (see Table~\ref{datatable}).  It is possible that LSBs exhibited episodic star formation and feedback at high redshift ($z \sim 2 - 4$) that could alter their halos as suggested by \citetalias{Pontzen2011}, but that would not leave a detectable signature in their stellar populations. However, one still needs a mechanism to ``shut off" this activity and replenish the pristine unevolved gas that is unique to LSBs.

\section{Summary and Outlook}
\label{discussion}
Detailed spectroscopic and photometric observations of LSBs indicate that their inner dark matter halos are round and cored. In order to reconcile these properties with the triaxial, cuspy halos produced in collisionless simulations of halo assembly, mechanisms that sphericalize and decrease the central density of the halo must operate.  

In this \textit{Letter}, we have compiled observed kinematics, structure, and star formation histories of a representative sample of late-type, bulgeless LSBs to examine the feasibility of baryons altering their halos. Invoking the currently favored methods for sphericalizing and erasing cusps in dwarf galaxies implies that LSBs must have had more baryons in the past that were then blown out by feedback from efficient star formation. We have argued that this is an unlikely scenario given the well-established observational properties of LSBs.

The properties of our LSB sample and state-of-the-art simulated ``bulgeless" galaxies are illustrated in Figure~\ref{dataplot}.  The dwarf galaxy of \citetalias{Governato2010} is similar to the observed LSBs of comparable mass, suggesting that if star formation is initiated, feedback and baryon removal can been effective at changing both the shape and density structure of the dark matter halo.  There are stark differences at high mass, however, between the LSBs and simulated galaxies.  Thus, while advances in simulating realistic galaxies have been made, no simulations yet resemble LSBs (nor claim to).  Because the simulated massive galaxies retain or re-accrete the baryons ejected during star formation feedback, their central halo densities are higher than both the densities of LSBs of similar mass and expectations for $\Lambda$CDM.  This is most likely due to adiabatic contraction during the formation of the simulated galaxies.   

What may be necessary for simulating LSBs with a wide range of masses is the combination of high(er) gas densities and low star formation efficiencies.  Of the most  recent simulations, \citetalias{Governato2010} impose the highest gas density required for star formation, $n =$ 100 atoms cm$^{-3}$, but this is just at the lower limit of the densities of giant molecular clouds.  Restricting star formation to only the highest density regions (which are also physically small) would be consistent with very little molecular gas being detected in LSBs \citep[e.g.][]{Das2006}, as the filling factor of these high density clouds would be low in a single resolution element of the observations.  But because regions with high gas densities are physically smaller than regions with low gas densities, increasing the density threshold for star formation requires very high numerical resolution, making this a computationally expensive endeavor.  

We reiterate here that while it is tempting to sweep LSBs into a ``special" class of galaxies that form in a biased subset of halos \citep[e.g.][]{Maccio2007}, LSBs are too numerous to be the tail-end of the galaxy distribution \citep[e.g.,][]{McGaugh1995}.   Until numerical simulations can resolve the scales on which star formation is taking place and can implement a complete treatment of ISM physics allowing the properties of LSBs to be reproduced, these galaxies remain a problem for the $\Lambda$CDM picture. \\

\acknowledgements
We thank Fabio Governato and Stacy McGaugh for helpful comments on an early version of this manuscript, and acknowledge funding from the Natural Sciences and Engineering Research Council of Canada (NSERC).  



\clearpage
\begin{landscape}
\begin{deluxetable}{lccccccccccl}
\tabletypesize{\footnotesize}
\tablecaption{Late-type LSB Galaxy Properties}
\tablewidth{0pt}
\tablehead{
\colhead{Galaxy} &\colhead{$V_{flat}$}  &\colhead{$\eta$} &\colhead{$\Delta_{V/2}$} &\colhead{$M_{total}$} &\colhead{$M_{HI}$}  &\colhead{$(B-V)$} &\colhead{$M_{\star}$} &\colhead{12 + log(O/H)} &\colhead{SFR} &\colhead{$<$SFR$>$$_{past}$} &\colhead{References}\\
\colhead{}  &\colhead{(km s$^{-1}$)} &\colhead{} &\colhead{(10$^{5}$)} &\colhead{(10$^{10}$ $M_{\sun}$)} &\colhead{(10$^{10}$ $M_{\sun}$)} &\colhead{(mag)} &\colhead{(10$^{10}$ $M_{\sun}$)} &\colhead{} &\colhead{$M_{\sun}$ yr$^{-1}$} &\colhead{$M_{\sun}$ yr$^{-1}$} &\colhead{}\\
\colhead{(1)}  &\colhead{(2)} &\colhead{(3)} &\colhead{(4)} &\colhead{(5)} &\colhead{(6)} &\colhead{(7)} &\colhead{(8)} &\colhead{(9)} &\colhead{(10)} &\colhead{(11)}  &\colhead{(12)}
}
\startdata
F568-1		&142 	&0.41	&4.5		&5.7		&0.50	&0.58	&0.31	&8.62	&0.074\footnotemark[a], 0.28\footnotemark[b]	&0.52\footnotemark[a]	&D01;K07;S09;vdH00;W09\\
UGC 4325	&123	&0.38	&5.9		&1.6		&0.10	&0.47	&0.10	&8.33	&0.058\footnotemark[a]					&0.072\footnotemark[a]	&K08;P07;S09;vZ00;vZ01\\
F568-3		&120	&0.48	&1.8		&5.6		&0.39	&0.61	&0.41	&8.68	&0.35\footnotemark[b]					&\nodata				&K08;M05;S09;vdH00;W09\\
F568-V1		&118	&0.26	&4.3		&5.6		&0.34	&0.57	&0.24	&8.61	&0.29\footnotemark[b]					&0.135\footnotemark[a]	&D01;K07;S09;vdH00;W09\\
F563-V2		&113	&0.45	&3.9		&2.6		&0.32	&0.51	&0.25	&8.10	&0.18\footnotemark[a]					&0.13\footnotemark[a]	&H05;K07;K04;K08;S09\\
F579-V1		&114	&0.46	&20		&4.3		&0.11	&0.70	&0.53	&\nodata	&0.27\footnotemark[b]					&\nodata				&B00;D01;D96;S09;W09\\
F563-1		&111	&0.38	&3.1		&5.2		&0.39	&0.64	&0.20	&8.02	&0.17\footnotemark[b]					&\nodata				&K04;K08;S09;W09\\
UGC 1230	&103	&0.41	&2.6		&8.5		&0.81	&0.54	&0.30	&7.91	&12.3\footnotemark[a]					&0.28\footnotemark[a]	&D02;K07;K04;S09\\
F574-1		&100	&0.43	&3.1		&2.9		&0.49	&0.49	&0.29	&\nodata	&0.33\footnotemark[b]					&\nodata				&B00;D01;S09;W09\\
UGC 5005	&99		&0.50	&0.67	&6.2		&0.41	&\nodata	&\nodata	&8.04	&\nodata								&\nodata				&D02;D98;K04\\
F583-1		&86		&0.30	&1.8		&2.5		&0.16	&0.39	&0.025	&\nodata	&0.055\footnotemark[a], 0.087\footnotemark[b]	&\nodata				&H05;K08;S09;W09\\
UGC 3371	&86		&0.38	&1.2		&1.8		&0.12	&0.72	&0.18	&8.48	&0.092\footnotemark[a]					&0.233\footnotemark[a]	&D02;H99;S09;vZ00;vZ01\\
NGC 4395	&83		&0.47	&12		&1.3		&0.20	&\nodata	&\nodata	&8.27	&\nodata								&\nodata				&K08;P04\\
UGC 5750	&79		&0.45	&0.39	&3.1		&0.14	&\nodata	&\nodata	&\nodata	&0.27\footnotemark[b]					&\nodata				&D98;K08;W09\\
NGC 3274	&79		&0.49	&39		&1.0		&0.085	&\nodata	&0.10	&8.33	&\nodata								&\nodata				&D02;H99;S09\\
NGC 1560	&78		&0.57	&6.0		&1.2		&0.12	&0.57	&0.050	&8.02	&\nodata								&\nodata				&D02;P04;S09\\
UGC 731		&75		&0.36	&15		&0.91	&0.074	&0.34	&0.18	&8.47	&0.16\footnotemark[a]					&0.041\footnotemark[a]	&D02;H99;K07;S09;S10\\
F583-4		&70		&0.38	&4.9		&0.76	&0.077	&\nodata	&\nodata	&\nodata	&\nodata								&\nodata				&D98;K08\\
NGC 4455	&64		&0.62	&2.4		&0.56	&0.061	&0.10	&\nodata	&\nodata	&0.37\footnotemark[a]					&6.61\footnotemark[a]	&D02;K07;S10\\
DDO 189		&64		&0.37	&4.1		&0.78	&0.13	&0.33	&0.010	&\nodata	&0.013\footnotemark[a]					&0.011\footnotemark[a]	&D02;S09;vZ00;vZ01\\
UGC 4173	&57		&0.62	&0.41	&0.91	&0.24	&\nodata	&\nodata	&\nodata	&\nodata								&\nodata				&D02;S10	\\
NGC 2366	&55		&0.62	&4.7		&0.38	&0.093	&\nodata	&0.13	&7.92	&0.095\footnotemark[a]					&\nodata				&D02;H05;P04;S09\\
DDO 185		&50		&0.34	&1.1		&0.13	&0.013	&0.43	&\nodata	&7.70	&0.14\footnotemark[a]					&0.009\footnotemark[a]	&D02;K07;L07
\enddata
\footnotetext[a]{H$\alpha$}
\footnotetext[b]{$UV$}
\tablecomments{Col.~(2): Flat rotational velocity of the galaxy.   Col.~(3): $\eta = V_{disk}/V_{circ}$ at $r = 2.2 R_{disk}$, calculated using the constant, non-zero $\Upsilon_{*}$ in the published mass models; see references.  Col.~(4): $\Delta_{V/2}$ parameter of \citet{Alam2002}.  Col.~(5): Total dynamical mass calculated at the last rotation curve point.  Col.~(6): Total \HI\ mass.  Col.~(7): Optical color.  Col.~(8): Stellar mass calculated using the $\Upsilon_{\star} - (B-V)$ relation of \citet{Stark09}.  Col.~(9): Mean oxygen abundance.  Col.~(10): Current star formation rate.  Col.~(11): Average past star formation rate.} 
\label{datatable}
\end{deluxetable}
\clearpage
\end{landscape}

\end{document}